\documentclass[12pt]{article}

\usepackage{scicite}
\usepackage{ragged2e}
\usepackage{times}
\usepackage{graphicx}
\graphicspath{ {./Images/} }
\usepackage{gensymb}
\usepackage{tabu}
\usepackage{enumitem}
\usepackage{amsmath}
\usepackage{textcomp}
\usepackage{caption}
\usepackage[T1]{fontenc}

\usepackage{times}

\newenvironment{sciabstract}{%
\begin{quote} \bf}
{\end{quote}}

\newcounter{lastnote}

\usepackage[compact]{titlesec}
\titlespacing{\section}{0pt}{*0}{*0}
\titlespacing{\subsection}{0pt}{*0}{*0}
\titlespacing{\subsubsection}{0pt}{*0}{*0}

\title{An interlayer with low solubility for lithium enhances tolerance to dendrite growth in solid state electrolytes} 

\author
{Vikalp Raj, Varun R Kankanallu, Bibhatsu Kuiri, Naga Phani B Aetukuri$^{\ast}$\\
	\\
	\normalsize{Solid State and Structural Chemistry Unit, Indian Institute of Science,}\\
	\normalsize{Bangalore, 560012, Karnataka, India}\\
	\\
	\normalsize{$^\ast$Corresponding Author; E-mail:  phani@iisc.ac.in}
}

\date{}

\begin{document} 


\baselineskip 24pt


\maketitle 


\begin{sciabstract}
  {All solid state Li-ion batteries employing metallic lithium as an anode offer higher energy densities while also being safer than conventional liquid electrolyte based Li-ion batteries. However, the growth of tiny filaments of lithium (dendrites) across the solid state electrolyte layer leads to premature shorting of cells and limits their practical viability. The microscopic mechanisms that lead to lithium dendrite growth in solid state cells are still unclear. Using garnet based lithium ion conductor as a model solid state electrolyte, we show that interfacial void growth during lithium dissolution precedes dendrite nucleation and growth. Using a simple electrostatic model, we show that current density at the edges of the voids could be amplified by as much as four orders of magnitude making the cells highly susceptible to dendrite growth after void formation. We propose the use of metallic interlayers with low solubility and high nucleation overpotential for lithium to delay void growth. These interlayers increase dendrite growth tolerance in solid state electrolytes without the undue necessity for high stack pressures.}
\end{sciabstract}


\section{Introduction}

\par Conventional liquid electrolyte-based Li-ion batteries (LIBs) led to the ubiquitous adoption of portable electronic devices. Such widespread adoption of all-electric vehicles requires batteries that offer long cycle and calendar life, and have a high energy density while also being safe when used in large battery packs\cite{Armand2008,Goodenough2013,Liu2019}. Solid-state Li-ion batteries that employ solid-state electrolytes (SSEs) in conjunction with a metallic lithium negative electrode are expected to meet the techno-economic attributes necessary for the electrification of transportation \cite{Manthiram2017,LiPON_Solid_electrolyte,albertus2017}. However, the rechargeability of such batteries relies critically on the uniform stripping and plating of lithium at the negative electrode during discharge and charge, respectively\cite{Janek2016,Famprikis2019,Zheng2014}. Non-uniform deposition of lithium leading to dendrite growth at the negative electrode has been observed in Li-ion cells employing liquid electrolytes\cite{Choi2016,Aetukuri2015}. It is expected that SSEs have sufficiently high shear moduli to suppress lithium dendrite growth and enable uniform lithium deposition\cite{Newman}. In stark contrast to this, there is growing evidence that the deposition of lithium at the negative electrode in cells employing SSEs leads to dendrite growth and subsequent cell shorting\cite{Critical_stripping_current_lead_to_dendrites,Cheng2016,Swamy_et_al_2018_simulation,electronic_conductivity,Tsai2016,SUDO2014151,C8EE00540K,Interface_stability_Jeff_temperature_CCD}. Surprisingly, cell shorting in solid state batteries was found to occur at much smaller current densities ($\sim$200 \micro A/cm$^2$) when compared to cells employing liquid electrolytes\cite{Cheng2016,formation_and_stability_of_interface,MANALASTAS2019287,Sharafi2017,Yu2019,Li2015}.

\par A mechanistic understanding of the growth of dendrites and possible solutions to mitigate dendrite growth in SSEs is indispensable for solid state battery development. Over the past few years several mechanisms for the possible origin of dendrite growth in SSEs have been proposed. For example, it was shown that the dendrite growth resistance in cells employing lithium phosphorous oxynitride (LiPON) based solid electrolytes is higher than cells employing Li${_{7}}$La${_{3}}$Zr${_{2}}$O${_{12}}$(LLZO) or Li${_{10}}$GeP${_{2}}$S${_{12}}$(LGPS) based SSEs. It is argued that nucleation of lithium within the SSE is possible in LLZO and LGPS due to their higher electronic conductivity, eventually leading to cell shorting\cite{electronic_conductivity}. In a different study, it was shown that dendrite growth selectively occurs through flaws in the SSE. Using an electrochemomechanical model, the authors suggested that the hydrostatic stress applied by a growing lithium filament at a crack tip in SSE leads to dendrite growth and subsequent crack propagation\cite{MECHANISM_OF_LI_METAL_PENETRATION}. Seemingly, bulk properties of the SSE such as its electronic conductivity and fracture toughness ultimately determine the utility of a SSE for developing solid state batteries (SSBs). The reliance on bulk properties restricts the choice of SSE materials available for the development of SSBs and may ultimately deter their development.

\par A possible approach to mitigate dendrite growth at high current densities is to cycle lithium under high pressures (commonly referred to as stack pressure). In the absence of an external pressure, the formation of voids, via lithium vacancy accumulation, at the interface between a lithium electrode and SSE is expected during lithium dissolution\cite{Towards_a_fundamental}. Voids could then lead to local current density concentration (hotspots) and a high local hydrostatic stress resulting in an electrochemomechanically driven dendrite growth and nucleation \cite{MECHANISM_OF_LI_METAL_PENETRATION,MANALASTAS2019287}. Consistent with this argument, uniform lithium plating was possible at high current densities in cells where void formation was suppressed by cycling cells under high stack pressures \cite{Towards_a_fundamental,Critical_stripping_current_lead_to_dendrites, wang2019}. These experiments are strong evidence that improvements to electrode-electrolyte interface can lead to enhanced dendrite growth resistance in SSEs with otherwise poor resistance to dendrite growth. However, extremely large pressures in the vicinity of 10-40 MPa are required for the suppression of void formation and dendrite growth \cite{Towards_a_fundamental}. While such large pressures lead to uniform lithium plating, they are impractical for battery development.
 
\par An alternate approach is to use thin metallic interlayers between the lithium electrode and SSE interface\cite{Lu2018,WangZnO_2017,alumina_coating_nature,Al_fu_2017,Solid_electrolyte_interfacial_challenges}. Metallic interlayers decrease interfacial impedance leading to an improved critical current density for lithium cycling without the necessity for large stack pressures. In cells without interlayers, we hypothesize that discontinuities may exist at the lithium-SSE interface as shown schematically in Fig. 1a. SSEs are poor electronic conductors with ionic conductivities that are orders of magnitude lower than electronic conductivities of metals. Therefore, any discontinuity at the lithium-SSE interface could lead to current density concentration at the edges of the discontinuities, finally resulting in dendrite growth. By contrast, a continuous interface is expected in cells with a metallic interlayer. Consequently, uniform current density distribution is expected at such interfaces, possibly increasing the average critical current densities for lithium cycling (Fig. 1b). Surprisingly, even in cells employing metallic interlayers, dendrite growth was observed at current densities as low as 200-300 \micro A/cm$^2$ \cite{Lu2018}.

\par The microscopic origins of dendrite growth in solid state cells with interlayers and possible ways to mitigate dendrite growth is critical for the practical realization of all solid state batteries (SSBs). Therefore, in this work, we chose to investigate the role of physical properties of the interlayer and the role of interfacial impedance on the dendrite-growth resistance of SSEs. We show that the dendrite growth resistance of SSEs can be deterministically controlled by altering the interfacial properties: interlayer material and/or interfacial impedance. We show that a seemingly bulk property of the electrolyte is extremely sensitive to interface optimization. Specifically, we propose that the overpotential for nucleation of lithium on the interlayer and the solubility of lithium in the interlayer are critical descriptors for choosing an ideal interlayer for cycling lithium at high current densities in all solid state cells. 
\vspace{5mm}
\section{Results and Discussion}
\begin{figure}[!htb]
\includegraphics[width=1\textwidth]{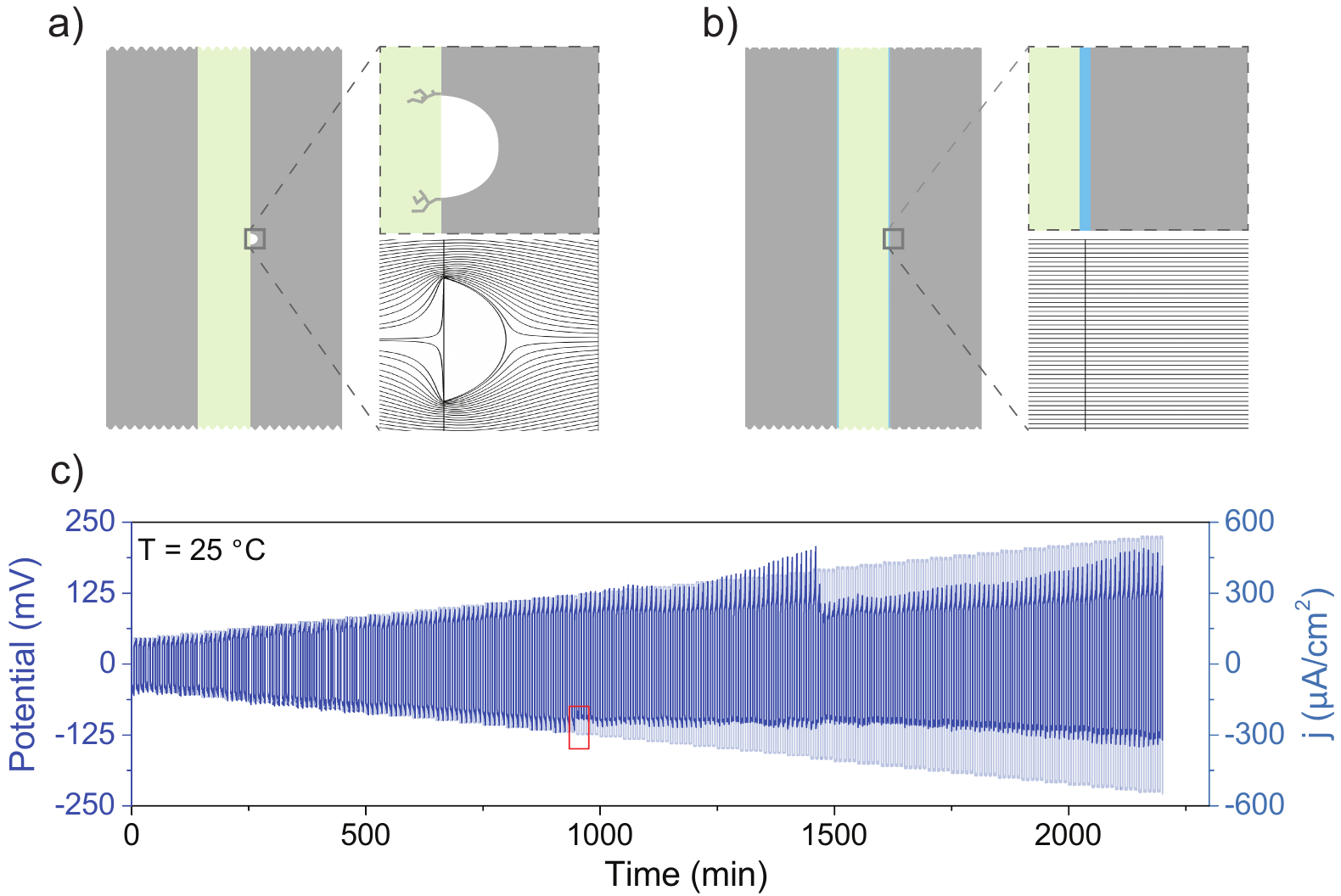}
\caption*{{\textbf{Fig. 1 | Cells with aluminum interlayer.} \textbf{a)} A schematic showing a Li/SSE/Li symmetric cell. In such cells, poor wetting of lithium with SSE can lead to areas with discontinuous interfacial contact between Li and SSE, resulting in current density concentration at void edges. \textbf{b)} A schematic showing a Li/Al/SSE/Al/Li symmetric cell. In cells with thin interlayers between SSE and lithium, better wettability of lithium with the interlayer leads to a continuous interfacial contact, resulting in a uniform current density distribution. \textbf{c)} A typical potential and current density versus time plot obtained from a critical current density experiment performed at a temperature of 25 $^{\circ}$C for a symmetric Li/Al/SSE/Al/Li cell. The red rectangular box is used to highlight a sudden drop in potential during the experiment. Such drops in potentials are identified to be due to electrical shorts caused by dendrite growth.}}
\end{figure}
\par First, we set out to study the role of interfacial impedance on the dendrite growth propensity of Li-Li half cells. For the studies reported in this section, aluminum (Al) \cite{Al_fu_2017,Lu2018} is the preferred interlayer and Li${_{6.4}}$La${_{3}}$Zr${_{1.4}}$Ta${_{0.6}}$O${_{12}}$(LLZTO) is the preferred SSE\cite{Murugan_LLZO_2007,C4CS00020J}. LLZTO was synthesized by solid state synthetic techniques that are based on a previously reported procedure\cite{Optimizing_LLZTO}. About 50 nm of Al was sputter deposited on sintered and polished LLZTO pellets which are approximately 12 mm in diameter. Lithium was then pressed onto the Al coated LLZTO pellets in an argon-filled glove box. The Li/Al/LLZTO/Al/Li symmetric cells were assembled in a custom-built flange-cell comprising of a low spring constant spring ensuring continual electrical contact. The maximum stack pressure due to this spring was calculated to be less than 160 kPa at all times. These symmetric cells were subjected to a galvanostatic training step before performing critical current density experiments. In the training step, a constant current density of 100 \micro A/cm$^2$ is applied cyclically (alternate lithium plating and stripping cycles) for 24 hours with 5 minutes each for a plate and strip cycle. After the galvanostatic training step, the cell is cycled at increasing current densities from 100 \micro A/cm$^2$ in steps of 10 \micro A/cm$^2$ to current densities beyond the critical current density. At each current density a cell is cycled between a 5-minute plate and 5-minute strip step for 5 times. The first instance of a partial but sudden drop in potential is considered as a short and the current density when such a drop in potential occurs is taken as the critical current density for dendrite growth (Fig. 1c). For further experimental details please see SI section 1.1-1.5, SI Figs. S1-S10 and SI Table. S1.

A typical potential versus time plot for a galvanostatic critical current density (CCD) experiment carried out at 25 \degree C is shown in Fig. 1c. In this experiment, a short occurred at a current density of 300 \micro A/cm$^2$. After a short, the lithium plating/stripping potential does not increase with increasing current density as expected from Ohm's law. In addition, impedance measurements after a short showed a clear decrease in impedance (Fig. S11, S12). Furthermore, we also observed a change in the activation energy for Li-ion conduction in SSE (measured by temperature dependent impedance spectroscopy) after shorting (Fig. S12). And in all cases, a short is preceded by an increase in the root-mean square (RMS) noise of the cell potential during a galvanostatic experiment (Fig. S13). We used these signatures of a short to precisely arrive at critical current densities for dendrite growth in all the data presented in this work \cite{albertus2017}.

\par The critical densities across several, nominally identical cells, varied between 200 to 400 \micro A/cm$^2$. While the range of current densities is consistent with previous reports \cite{Lu2018}, the origin of such a wide variation in critical current density is not clear. As mentioned earlier and in SI section 1.1-1.3, cells were prepared from SSE pellets that were polished manually and then lithium was pressed onto the Al-coated pellets using a thermal diffusive approach. We hypothesized that one or a combination of these approaches results in a natural variation in the interfacial impedance of the cells leading to variable critical current densities. In Fig. 2a, we plot critical current density as a function of interfacial impedance. Interfacial impedance is calculated using an approach that was previously reported\cite{Al_fu_2017} (see SI section 1.6 and SI Fig. S4). Clearly, interfacial impedance and critical current density are inversely correlated. This suggests that better interfacial contact is necessary for a high critical current density. However, the variation in interfacial impedance is not very well controlled and the evidence that poor interfacial contact leads to dendrite growth is at best indirect. Therefore, we introduced a controlled discontinuity in the interlayer to deterministically probe the role of interfacial impedance as discussed below.
\begin{figure*}[!htb]
	\includegraphics[width=1\textwidth]{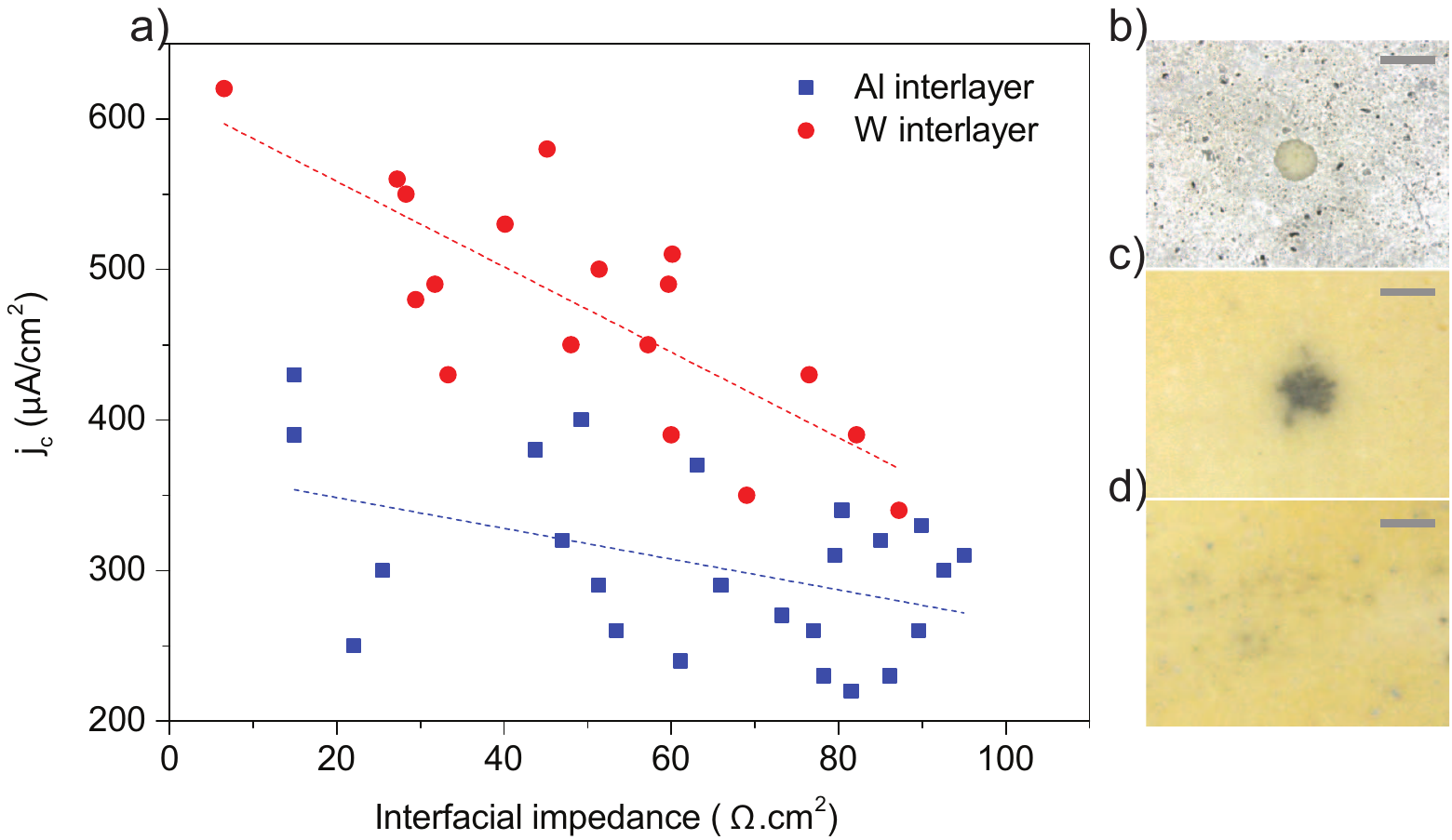}
	\caption*{{\textbf{Fig. 2 | Critical Current Density and Interfacial Impedance.} \textbf{a)} A comparison of critical current density vs area specific interfacial impedance for symmetric Li/Al/LLZTO/Al/Li and Li/W/LLZTO/W/Li cells. Generally, cells with low interfacial impedance show a higher stability towards dendrite growth. Amongst cells with similar interfacial impedance, cells with W interlayers have a higher current density threshold for dendrite growth in comparison to cells with Al interlayers. \textbf{b)} Optical microscopy image of a pellet with an aluminum interlayer deposited across the entire surface except for a tiny patterned region in the center of the pellet; \textbf{c)} Optical microscopy image of the same pellet as in \textbf{b} but after the cell is cycled until it is shorted. The dark spot in the center is due to dendrite growth in the region without the interlayer. \textbf{d)} Optical microscopy image of the same pellet but on the other side of the pellet where the aluminum interlayer completely covers the surface. Signatures of dendrite growth are not apparent. Optical images in \textbf{c} and \textbf{d} were taken after the pellet surfaces were mechanically polished. The scale bars in \textbf{b}, \textbf{c},  \textbf{d} are all equivalent to 1 mm.}}
\end{figure*}
\par To break the symmetric half cell structure, we deposited aluminum interlayer on one side of the SSE pellet. On the other side, we used patterning techniques to leave a small region of SSE without any aluminum interlayer. The area of the region without the aluminum interlayer is less than 1\% of the total area ($\sim$ 1 mm in diameter compared to a pellet diameter of 10 mm) (see Fig. 2a). Therefore, the interfacial impedance does not change significantly. Consistent with this, the critical current density, measured from galvanostatic experiments, does not change. However, upon visual inspection, we observed black spots on the pellet after the cell shorted. These black spots were always seen in pellets after a galvanostatic experiment  was run until the cells showed a short (Figs. 2b-d and SI Fig. S14). Such spots were not seen in cells cycled to current densities below their critical current densities (SI Fig. S15). Therefore, we believe that dendrite growth was prominent at these sites with black spots\cite{formation_and_stability_of_interface}. Interestingly, the spots were seen consistently in the area without the interlayer in all the cells that we have tested in this configuration. It is important to note that the region without the interlayer has a high ionic interfacial impedance (SI Fig. S9 and S10). Hence, the current density through the region is expected to be smaller than the average current density. It is rather surprising that a region that contributes a smaller portion of the total current through the cell is the region where the dendrite growth is concentrated. 

\par In addition to where the SSE pellet is patterned, we also observed signatures of dendrite growth at other areas on the pellet (SI Fig. S14). But the location of these spots are randomly distributed. We hypothesize that in cells with symmetrically deposited interlayers, there are either inherent defects on SSE surfaces that lead to high local impedance or that such discontinuities develop during the course of cycling or a combination of two. In general, the strong anti-correlation between interfacial impedance and critical current density could be attributed to the defects at the SSE interface that are present even before cell cycling. In addition to this, defects could form dynamically at the interface during cell cycling. For example, micron-scale void formation is expected during lithium plating and stripping in cells cycled under low stack pressures \cite{Critical_stripping_current_lead_to_dendrites}. Although, much smaller in size compared to the patterned defect, voids are regions that are electrically discontinuous. Therefore, voids can be considered as scaled down versions of the patterned defect discussed above. Scanning electron microscopy (SEM) images performed on cross-sections of cycled cells showed the presence of voids at lithium/Al/SSE interfaces (Fig. 3a). This is evidence that void formation occurs at the interface of lithium and SSE in cells employing Al interlayers. Clearly, even when an ideal interface is formed during cell fabrication, void growth may lead to local discontinuities and subsequent dendrite growth. A natural question to ask is, why would voids lead to dendrite growth and how can dendrite growth be mitigated?
\begin{figure*}[!htb]
	\includegraphics[width=1\textwidth]{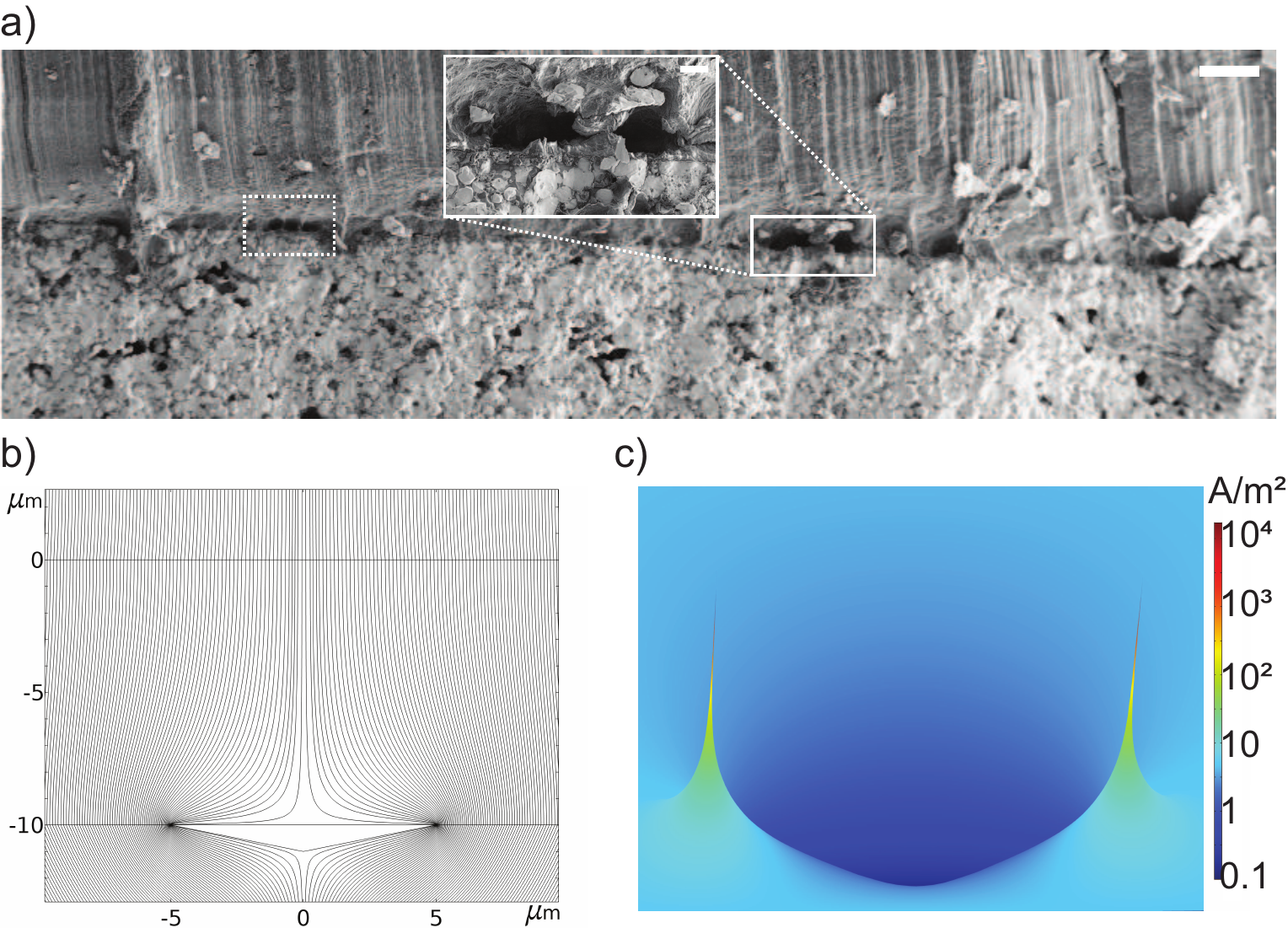}
	\caption*{{\textbf{Fig. 3 | Interfacial voids as precursors for dendrite growth.} \textbf{a)} A cross-sectional SEM image at Li/Al/LLZTO interface of a shorted cell. Interfacial voids can be clearly seen. Inset shows a magnified image of the voids. The scale bars in \textbf{a} and inset of \textbf{a} are equivalent to 100 \micro m and 20 \micro m, respectively. \textbf{b)} A simulated current density contour map in the vicinity of a void showing that the constant current density lines bend around the void center and concentrate near the edges of the void. \textbf{c)} A 3D color map of the same data as in \textbf{b} showing that the current density concentration at the edges is amplified by four orders of magnitude to 10$^4$ A/m$^2$. Simulations were performed for an average current density of 5 A/m$^2$.}}
\end{figure*}
\par Intuitively, since the void does not conduct, the current detours around the void concentrating in a small region at the edges of the void where lithium contacts the solid-state electrolyte. The resultant local current density enhancement could lead to the formation of lithium filaments either via electromigration or by a process similar to the growth of filaments in resistive memories or via the electrochemomechanical mechanism as reported by others\cite{Yang2012,Onofrio2015,MECHANISM_OF_LI_METAL_PENETRATION}. Irrespective of the mechanism, as long as an interface between lithium and the solid-electrolyte exists such hotspots with high current density will be inevitable. 

\par To quantify this effect, we modeled current density distribution in the vicinity of a void in an otherwise ideal interface. For simulating current density distributions around the defect, we used the electric current module in COMSOL mutiphysics simulation package (see SI section 1.9 and SI Fig. S5 for details). Current density distribution obtained from the simulation is plotted in Fig. 3b,c. The simulations show that local current density at void edges could be as high as 1 A/cm$^2$ (10$^4$ A/m$^2$) for a cell current density of 0.5 mA/cm$^2$ (5 A/m$^2$). These results imply that dendrite growth may be occurring at local current densities that are orders of magnitude larger than the average current density in experimental cells. 

\par For practically viable SSBs, it is sufficient if cells can cycle at an average current density of 10 mA/cm$^2$\cite{albertus2017}, which is two orders of magnitude lower than the local current densities at which dendrite growth may be occurring. Clearly, it should be possible to develop practical solid state batteries without dendrite growth by developing strategies to avoid local current density concentration. This necessitates avoiding void growth during lithium dissolution. As stated earlier, large stack pressures could be used for arresting void growth at lithium/SSE interfaces, but this is impractical.

\par As an alternative, we propose the use of metallic interlayers with ultra low solubility for lithium for minimizing void growth at lithium/SSE interfaces. We note that most of the work so far has concentrated on interlayers that alloy with lithium \cite{Tsai2016,Al_fu_2017}. In general interlayers that can alloy with lithium have been preferred because alloying creates a lithium rich phase at the interface thus minimizing potential drop across the interlayer. However, after alloying there is no distinct interlayer present and lithium dissolution can occur from the alloyed interlayer leaving behind voids at the interface with SSE (Fig. 4a, b). Therefore, metallic interlayers with low solubility for lithium and high overpotential for lithium nucleation are possible candidates for decreasing the propensity for void growth. If the interlayer is sufficiently thin, it would be favorable for lithium to transport across the interlayer and deposit (or grow) on a pre-existing lithium film. This avoids any lithium nucleation at the interface with SSE. Note that this approach would work only if there is lithium already deposited on the interlayer. In addition, during dissolution, the low solubility of lithium in the interlayer ensures that there is not sufficient concentration of lithium to facilitate void growth by vacancy migration (Fig. 4c, d). Interestingly, metals with low solubility for lithium also have a high overpotential for lithium nucleation. Thus, using interlayers with low solubility for lithium could potentially avoid lithium nucleation at the interface between the interlayer and the SSE.
\begin{figure*}[!htb]
	\includegraphics[width=1\textwidth]{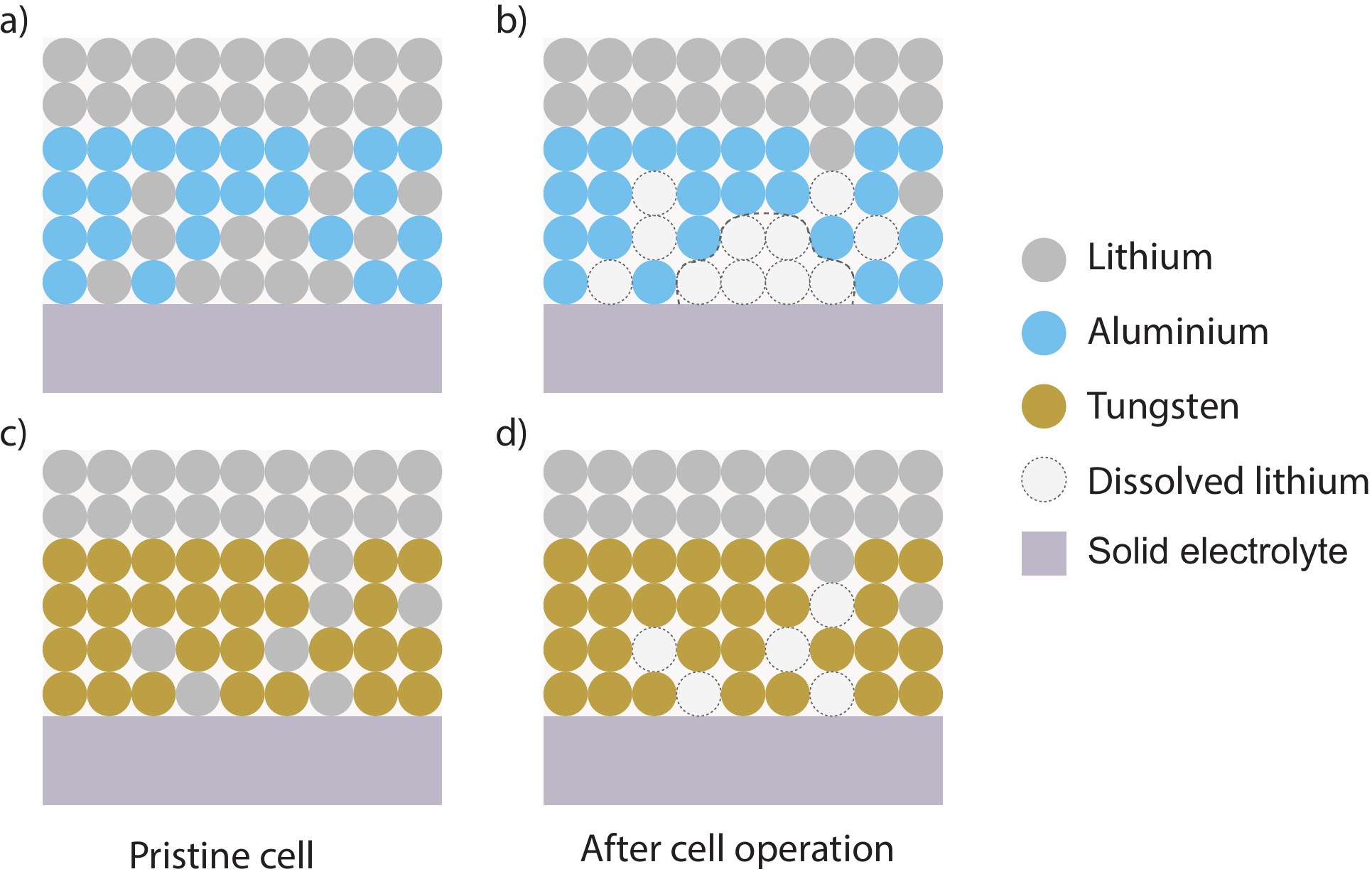}
	\caption*{{\textbf{Fig. 4 | Lithium dissolution in cells with interlayers}. Schematics comparing the solubility and lithium dissolution in cells with \textbf{a, b} an aluminum interlayer and \textbf{c, d} a tungsten interlayer. Due to the lower solubility of lithium in tungsten, vacancy migration driven void growth is expected to be suppressed in cells employing tungsten as an interlayer.}}
\end{figure*}
\par Tungsten(W) and related refractory metals have low solubility for lithium \cite{Li_W_NASA,Li_W_book,Sangster1991_li_w,De_mastry_li_w} and could therefore be candidate materials for minimizing void growth at the interface with SSE. We performed galvanostatic lithium plating at a constant current density of 100 \micro A/cm$^2$ on a 200 nm Al film deposited by DC magnetron sputtering and a 50 \micro m thin foil of W (see SI section 1.8 for details). For the deposition of Li on Al (Fig. 5a), there is a small over potential for lithium nucleation in the first cycle, but there is almost no difference between the plating potential and the nucleation potential for subsequent cycles. Note that lithium has been completely stripped after every plating step. Similar results were obtained for lithium plating on Al by others where the differences between the first cycle and subsequent cycles were attributed to alloying occurring after the first cycle. In complete contrast to the results for Al, galvanostatic plating of Li on W showed a clear dip in potential within the first 100 seconds of deposition for the first and subsequent cycles (Fig. 5b). The potential then increases and stabilizes during the rest of the plating process. This dip in potential at the start of the deposition has been identified to be the nucleation overpotential \cite{Yan2016}. Clearly, the nucleation overpotential is independent of the cycle number for Li nucleation on W. This suggests that no significant alloying has occurred during the plating process. Therefore, using W as an interlayer can potentially increase the critical current densities for dendrite growth in SSEs.
\begin{figure*}[!htb]
	\includegraphics[width=1\textwidth]{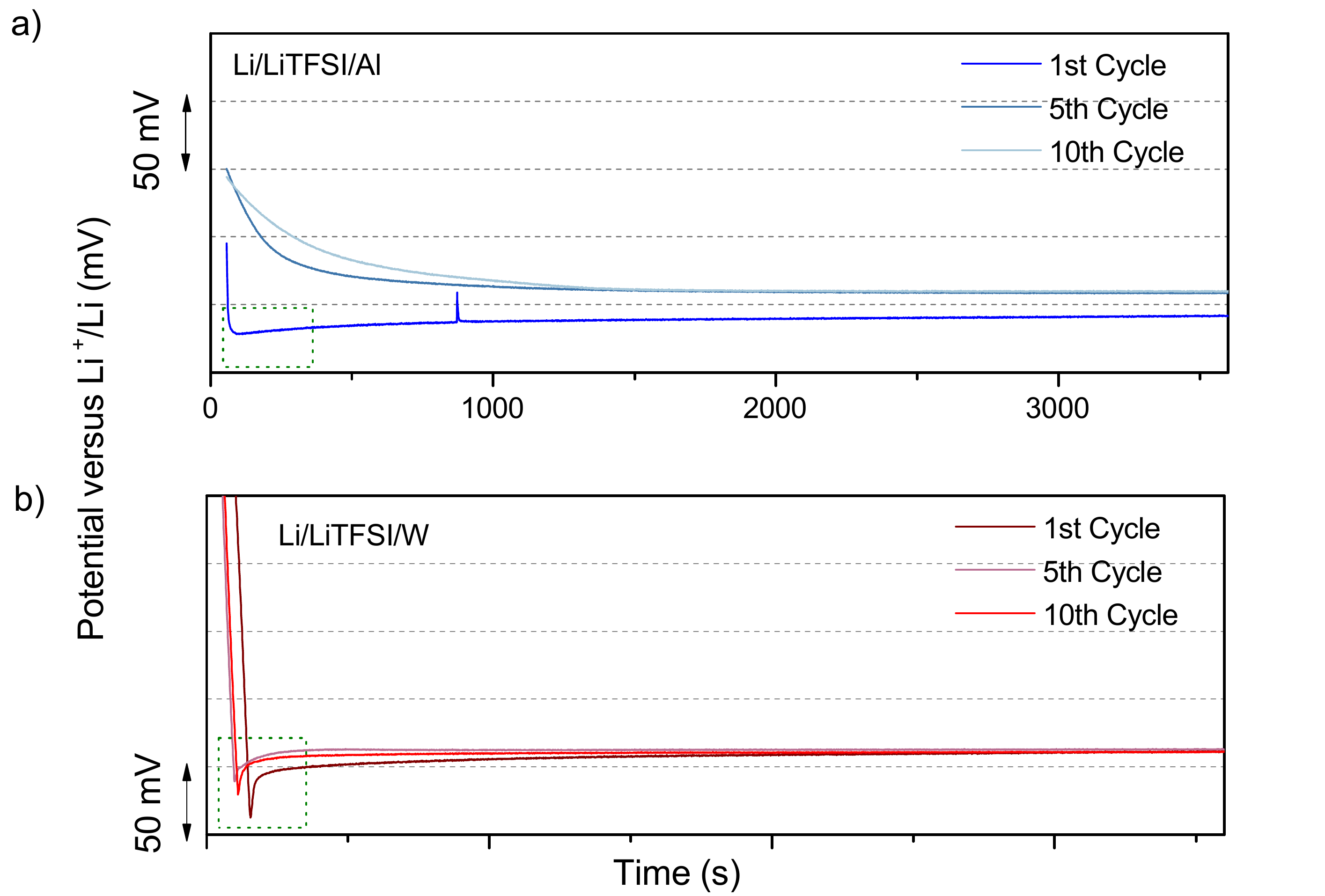}
	\caption*{{\textbf{Fig. 5 | Lithium nucleation overpotential for Al and W.} Potential versus time plot for galvanostatic lithium plating experiments on a) Al and b) W performed at 100 \micro A/cm$^2$. The nucleation overpotential for lithium on aluminum is non-existent after the first cycle. However, there is a clear dip in potential near about 100 seconds for all cycles when lithium is plated on tungsten.}}
\end{figure*}
\par In Fig. 6a, we show a typical potential and current density versus time plot for a galvanostatic critical current density experiment carried out at room temperature for a Li/W/LLZO/W/Li cell. All experiments were performed on cells with ~30 nm of tungsten. Clearly, the critical current density of 530 \micro A/cm$^2$ is much higher than the best cells with aluminum interlayers. Comparing across cells with similar interfacial resistance, cells with tungsten interlayers always lead to higher critical current densities in comparison to cells with aluminum interlayers (Fig. 2a). Furthermore, cells with tungsten interlayers had higher average critical current densities for dendrite growth for all temperatures at which experiments were performed (Fig. 6c). In fact, cells with tungsten interlayers have current densities approaching $\sim$2 mA/cm$^2$ at the highest temperature of 70 \degree C while aluminum cells show current densities that are approximately half this value. In addition, cells with tungsten interlayers could be cycled for at least 200 hours at a current density of 400 \micro A/cm$^2$ at room temperature (Fig. 6c). At 60 \degree C, the cells can be cycled without shorting at current densities as high as 1 mA/cm$^2$ for 200 hours (Fig. 6d). In comparison, the average critical current density for cells with aluminum interlayers is only $\sim$300 \micro A/cm$^2$ at room temperature and $\sim$750 \micro A/cm$^2$ at 60 \degree C (see SI Figs. S16 to S19 ). 
\begin{figure*}[!htb]
	\includegraphics[width=1.0\textwidth]{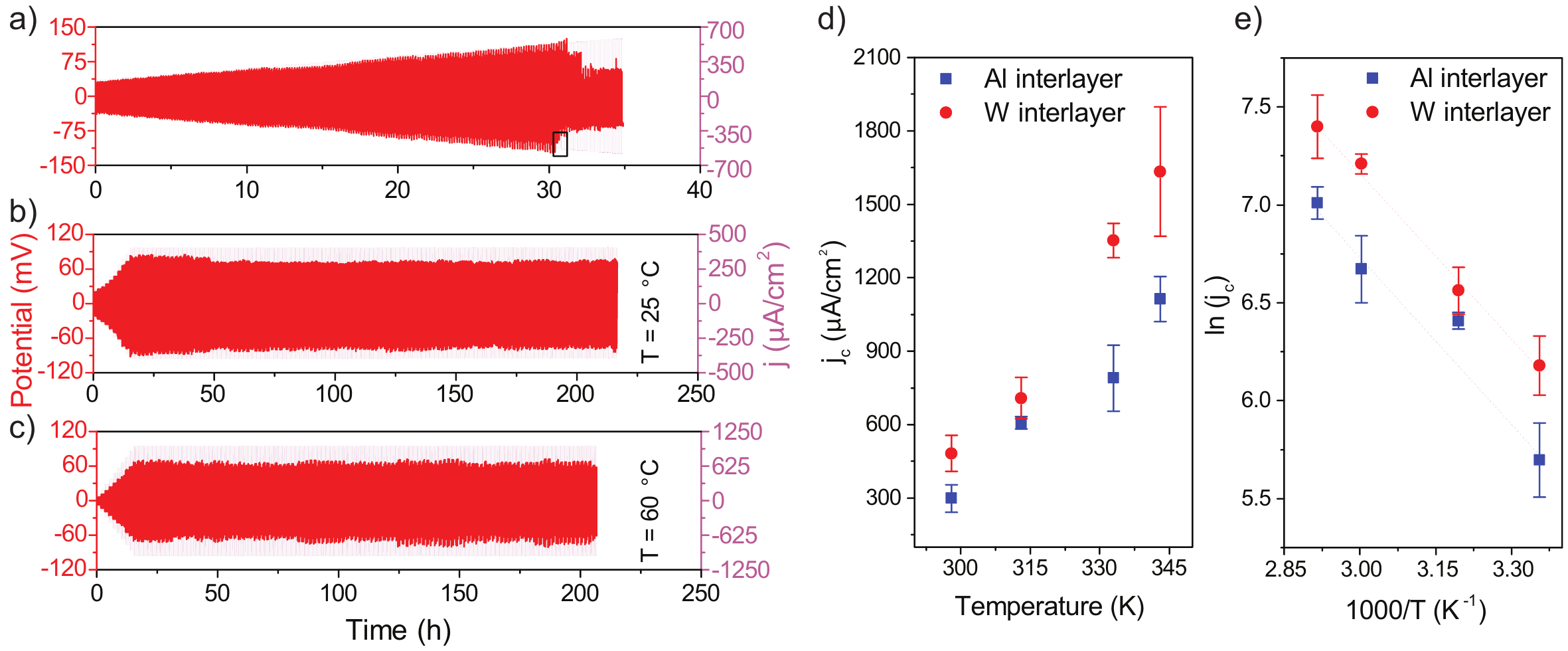}
	\caption*{{\textbf{Fig. 6 | Cells with tungsten interlayers.} \textbf{a)} A typical potential and current density versus time plot obtained from a critical current density experiment performed at a temperature of 25 $^{\circ}$C for a symmetric Li/W/SSE/W/Li cell. The black box is used to identify first instance of a potential drop signifying an electrical short. This corresponds to a current density of 530 \micro A/cm$^2$ which is considerably higher than the critical current density of Al interlayer cell 300 \micro A/cm$^2$ \textbf{(Fig. 1c)}. In figure \textbf{b)} and \textbf{c)} Galvanostatic cycling of a Li/W/SSE/W/Li cell at room temperature at a current density of 400 \micro A/cm$^2$ and 1 mA/cm$^2$ respectively. The cells are cycled to plate and strip 200 hours. \textbf{d)} A comparison of critical current density at four different temperatures of 25 $^{\circ}$C, 40 $^{\circ}$C, 60 $^{\circ}$C and 70 $^{\circ}$C for Li/Al/SSE/Al/Li and Li/W/SSE/W/Li cells. Clearly, the performance of cells with W interlayers continues to be better than the cells with Al interlayers at all temperatures used in this study. \textbf{e)} A plot of the logarithm of current density versus the reciprocal of temperature suggesting a nearly exponential increase in critical current density with temperature.}}
\end{figure*}
\par These results indicate that tungsten interlayers could enhance critical current densities for dendrite growth in solid state batteries without the need for high stack pressures. However, we note that the critical current densities achieved with tungsten are still at least a few times less than required for practical applications. Cross sectional SEM imaging of the interfaces after shorting in cells employing tungsten interlayers showed the presence of voids (Fig. S20). In cells with tungsten interlayers, we hypothesize that this could be because there are two competing processes: 1) Lithium deposition on a preexisting lithium film by transporting across the interlayer and 2) Lithium nucleation at the interface between the tungsten interlayer and the SSE. The former will be favored at low current densities while the latter will be favored at high current densities. This cross-over current density could be a function of the thickness of the interlayer. Consistent with this, we observed that cells with thicker tungsten interlayers shorted at lower current densities (see Figs. S21and S22). However, in cells with thinner tungsten interlayers, we found that tungsten was oxidizing (as tungsten deposition was done in a chamber not connected to the argon glove box) leading to increased initial polarization. Approaches that enable thinner tungsten interlayers could lead to further increase in critical current densities. 

\par Interestingly, cells with aluminum and tungsten interlayers both show a nearly exponential rise in the critical current density with increasing temperature (Fig. 6e). The slopes of $ln(j_c$) vs $1/T$ are nearly identical for both Al and W. This is suggestive of a common dendrite growth mechanism in both cells. Perhaps, lithium nucleation at the interface between the interlayer and SSE occurs at the critical current density, subsequently leading to void growth and current density hot spots. Notably, such an exponential increase in critical current density is proposed to signify dendrite growth originating from voids in cells without interlayers\cite{Towards_a_fundamental,wang2019}. 
\par We also observed critical current density enhancement in cells employing molybdenum as an interlayer (see SI Figs. S23 to S25) . Akin to tungsten, molybdenum has a low solubility for lithium. The observation of high critical current density for dendrite growth in cells employing Mo and W as interlayers might very well be indicative of a common mechanism that leads to enhanced resilience for dendrite growth in these cells. Since the SSE was not changed in our experiments, bulk properties of SSE such as the electronic conductivity or changes in the mechanical fracture toughness can be ruled out. Clearly, our results imply that interlayers with low solubility and/or high nucleation overpotential for lithium might be the key for practical realization of SSBs. Furthermore, interface engineering could still enhance the stability of solid state electrolytes against premature shorting due to dendrite growth. 

\vspace{5mm}
\section{Additional information}
All materials, methods, calculations and additional data are available in supplementary information.

\vspace{5mm}
\section{Acknowledgments}
We would like to thank Supriya A from the Center for Nano Science and Engineering (CeNSE) for help with RF sputtering, Prof. Abhishek Mondal and his research group for access to optical microscope, and Jibin J Samuel for discussions. We would also like to acknowledge access to common facilities at CeNSE and SSCU. This work was supported by a grant from the Indian Space Research Organization (ISRO) through grant No. ISTC/CSS/NPH/397. N.B.A acknowledges the new faculty start-up grant no. 12-0205-0618-77 provided by the Indian Institute of Science.

\vspace{5mm}
\bibliography{scibib}

\end{document}